# Investigation on Structural, Optical, Thermal, and Magnetic Properties of Bismuth Ferrite Nanoparticles Synthesized at Lower Annealing Temperature

*Naresh Prajapati[a], G. Surya Prakash[a], Manoj Kumar[b], Himanshu Pandey[a,*]*

[a]Condensed Matter Low-Dimensional Systems Laboratory, Department of Physics, Sardar Vallabhbhai National Institute of Technology Surat-395007, Gujarat, India

[b]Department of Physics and Materials Science and Engineering, Jaypee Institute of Information Technology, Noida-201309, India

*Corresponding Author's email: hp@phy.svnit.ac.in*

Due to its multiferroic properties and narrow optical bandgap, Bismuth ferrite ($BiFeO_3$) has been widely explored for spintronics, photovoltaics, and photocatalysis applications. Bismuth ferrite can be synthesized in various forms like bulk, thin films, and nanostructures using various synthesis techniques. It is challenging to synthesize the pure $BiFeO_3$ phase due to the volatile nature of bismuth and the very narrow temperature range for forming this phase. So, this work aims to synthesize the pure $BiFeO_3$ phase at lower annealing temperatures using an efficient sol-gel method. We have chosen the annealing temperature from 450°C to 650 °C, and a detailed analysis of structural and optical properties is performed here. X-ray diffraction is used to confirm the crystalline nature of the material. Single-phase Rietveld analysis of XRD patterns is carried out to study the effect of annealing temperature on structural parameters. All the samples are crystalized in pure rhombohedral $BiFeO_3$ phase with the *R3c* space group symmetry, except those annealed at higher temperatures, 600°C and 650°C. Strain and dislocation densities were decreasing with an increase in the annealing temperature. From the UV-visible analysis, a strong response is observed below 600 nm in the visible region, and the band gap from the absorption behaviour is estimated in the range of 2.26 - 2.60 eV for these Bismuth ferrite nanoparticles. Fourier transform infrared analysis confirmed the existence of metal-oxygen bonds in Bismuth ferrite nanoparticles. These nanoparticles were found to be thermally stable from the thermal analysis performed using differential scanning calorimetry. Bismuth ferrite nanoparticles were weakly magnetic from the vibrating sample magnetometry analysis.

**Keywords:** Nanoparticles; Multiferroic; Perovskite; Sol-gel method; Rietveld analysis.

## 1. Introduction

Multiferroic materials exhibit multiple ferroic properties, such as ferroelectricity, ferromagnetism, and ferroelasticity in the same phase [1]. They have been widely explored for their applications in multifunctional low-power electronic devices, energy storage devices, and other environment-friendly applications [1,2]. Researchers are also interested in these materials to investigate the coexistence of different order parameters in the crystalline phase [3]. The coupling between these various parameters has generated another physical property, the magnetoelectric coupling [4,5]. Many efforts have been made to obtain more than one ferroic order at room temperature in a single phase [6]. However, the problem is that many of these materials don't exhibit multiferroic behaviour at room temperature. Most of them are not usable in devices that work at ambient temperature. Though few materials exhibit multiferroic behaviour at room temperature, bismuth ferrite (BiFeO$_3$) is one of them [5]. BiFeO$_3$ (BFO), the most studied multiferroic, is a rhombohedrally distorted multiferroic perovskite material (space group: *The R3c*) with an *AB*O$_3$-type structure. Perovskite materials with this structure are widely studied for their electrical, magnetic, and optical properties [5,6]. The unit cell of BFO is represented in a hexagonal reference frame with unit cell parameters $a$ = 5.58 Å and $c$ = 13.87 Å. BFO shows remarkable ferroelectric behaviour with Curie temperature around 1103 *K* and *G-type* antiferromagnetic character with Néel temperature around 643 *K* [7,8]. The ferroelectric behaviour of BFO is due to the existence of a $6s^2$ lone pair in the Bi$^{+3}$ ion, which is situated at the *A*-site of the unit cell that results in non-zero spontaneous polarization. Similarly, the Fe$^{+3}$ ion at the *B*-site is also responsible for the magnetic properties in BFO [9,10].

The effect of annealing temperature on the crystal structure of BFO is responsible for the variation in structural, electrical, magnetic, and optical properties [10-12]. BFO possesses rhombohedral *R3c* symmetry at room temperature but shows the transition to the orthorhombic phase *Pnma* at higher temperatures [13]. Even the cubic phase is reported near the melting point of BFO [14]. Other than the annealing temperature, the volatile nature of Bismuth plays an essential role in the pure phase formation. At higher annealing temperatures, Bismuth evaporates and causes a variation in the stoichiometry, resulting in the formation of other phases. Studies have been conducted where BFO was deliberately synthesized with different stoichiometry [15], and other phases like Bi$_2$Fe$_4$O$_9$ and Bi$_{25}$FeO$_{40}$ have been formed. Therefore, optimization of annealing temperature with proper stoichiometry is crucial for forming a pure BiFeO$_3$ phase. Researchers worldwide have tried techniques like doping, forming solid solutions, *etc*. [15,16] to synthesize BFO without the impurity phase. It is important to note that slight variations in the structural properties can be beneficial in manipulating the properties of any material [12]. The structural purity of perovskite material is measured using the Goldschmidt tolerance factor (*t*). It is given by

$$t = \frac{(r_A + r_O)}{\sqrt{2}(r_B + r_O)}. \tag{1}$$

Where *r* is the respective ionic radius. For stable perovskite material, *t* is usually between 0.7 to 1 [17]. For an ideal cubic perovskite, the value of *t* is close to 1, whereas for distorted perovskites, it lies between 0.7-0.9. For BFO, the value of *t* is around 0.89, which puts BFO in the category of rhombohedrally distorted perovskites [18]. Therefore, it is essential to have detailed structural information on BFO to investigate other properties such as optical, magnetic, dielectric, *etc*.

BFO has been synthesized using different techniques such as solid-state reaction, hydrothermal, sol-gel, and co-precipitation to prepare its bulk, thin films, nanoparticles, and other nanostructures to investigate the effect on structure, morphology, particle size, and other properties that might be useful for intended applications of BFO [16,18,19-25]. During annealing, the formation of different phases causes changes in the various properties of BFO [12]. Many studies have reported the formation of these impurity phases at lower annealing temperatures, but these phases disappear at higher annealing temperatures. Particle size significantly increases at higher annealing temperatures, especially while synthesizing the BFO nanoparticles. Hence, an optimization of the annealing is much required for the controlled synthesis of pure BFO. Annealing temperatures around 650 °C as well as around 780 °C have been reported for the formation of pure BFO phase [12,25]. This study aims to check whether single-phase BFO nanoparticles can be synthesized at a lower annealing temperature using an efficient low-temperature sol-gel synthesis method. It is well known that the temperature range for the formation of pure phase is very narrow [27,28]. So, we start with a low annealing temperature of 450 °C and increase to 650 °C to study the variation in structural, vibrational, optical, and magnetic properties with annealing temperature.

## 2. Experimental Details

### 2.1. Synthesis of the material

BFO nanoparticles were synthesized using the tartaric acid-assisted low-temperature sol-gel method. For the preparation of BFO nanoparticles, we have used bismuth nitrate pentahydrate [(Bi(NO$_3$)$_3$·5H$_2$O), Sigma-Aldrich] and ferric nitrate nonahydrate [(Fe(NO$_3$)$_3$·9H$_2$O), Sigma-

Aldrich] as a precursor. A stoichiometric amount of nitrate salts was weighed and dissolved in distilled water separately. A few drops of concentrated nitric acid were added to the solution of bismuth nitrate to dissolve it completely. The solution was stirred till it became transparent. The stoichiometric amount of chelating agent tartaric acid [($C_4H_6O_6$), Sigma-Aldrich] was (1:1 molar proportion with nitrates) dissolved in distilled water. All solutions were mixed and then continuously stirred and heated at 60-70 °C for 2-3 hours until all solvent evaporated and the solution was converted into a thick viscous gel. The gel was then dried at 100-120 °C for 24 hours to transform into dry brown xerogel. This xerogel was ground in an agate mortar pestle for 3-4 hours and converted into a fine powder. This fine powder was divided into five nearly equal proportions and then annealed in the tubular PID-controlled furnace at different annealing temperatures between 450 °C and 650 °C for 2 hours. The heating rate of the furnace was kept at 5 °C min$^{-1}$. These BFO nanoparticles annealed at 450 °C, 500 °C, 550 °C, 600 °C and 650 °C were labelled as $T_1$, $T_2$, $T_3$, $T_4$, and $T_5$ and were further characterized for the study of structural, optical, vibrational, thermal, and magnetic properties using different characterization techniques.

## 2.2. Characterization of the material

The structural property of synthesized BFO nanoparticles was investigated by X-ray diffraction (XRD) using the Rigaku Smart lab SE diffractometer with Cu $K\alpha$ radiation ($\lambda = 0.15425$ nm) in the $2\theta$ range of 20°- 80° with the step size of 0.01°. Rietveld analysis of the XRD pattern was carried out using the FullProf suite to study the effect of temperature on various structural parameters like cell parameters, crystallite size, strain, bond lengths, and bond angles. Fourier Transform Infrared (FTIR) analysis was carried out for vibrational properties and identification of chemical bonds using JASCO FT/IR-6600 type A in the 400-4000 cm$^{-1}$ range. UV-visible spectroscopy is used to study the optical response, and Tauc's plot is employed to evaluate the bandgap from the transmittance behaviour. UV-Vis Spectrophotometer (Thermo Scientific Evolution 260, Thermo Fisher Scientific Inc. USA) was used to record the absorbance spectra in the UV-visible-near IR region ranging from 190 to 1100 nm. To understand the physical and chemical changes including changes in heat capacity or exothermic/endothermic processes in the synthesized nanoparticles, thermal analysis measurement was performed using differential scanning calorimetry (Hitachi, DSC7000X) in the temperature range of 40 - 400 °C with a heating rate of 10 °C. The room-temperature magnetic properties were studied using a vibrating sample magnetometer (VSM, Lakeshore, USA) in an applied external magnetic field range between ±10,000 gauss.

## 3. Results and Discussion

### 3.1. XRD analysis

The XRD profiles of BFO nanoparticles, as shown in Fig. 1(a), depict all the significant peaks of the rhombohedral phase with the *R3c* space group symmetry. All peaks have matched with standard data from JCPDS card no. 71-2494. Major diffraction peaks of BFO are observed at 22.43°, 31.75°, 32.10°, 37.66°, 38.97°, 45.78°, 51.34°, 56.99°, and 57.19° corresponding to (012), (104), (110), (113), (006), (024), (116), (214), and (300) Miller planes, respectively. For samples $T_4$ and $T_5$, peaks from other phases like $Bi_2Fe_4O_9$ and $Bi_{25}FeO_{40}$ are also observed as marked with * and #, respectively, in Fig. 1(a). These phases might have formed due to the deviation in the stoichiometry at higher annealing temperatures on account of the volatile nature of Bismuth [15]. The peak position represents the material's crystal structure and the space group symmetry to which it belongs. Our synthesized material belongs to the *R3c* group with a rhombohedrally distorted perovskite. Variation in the intensity of diffraction peaks with the annealing temperature is related to the dynamics of crystal growth. Shift in the doublet peaks (104) and (110) is observed with the increase in the annealing temperature. This has already been reported by other researchers, too [25,26]. For better visualization, the change in the positions of (012); and (104) and (110) Braggs peaks are shown by *2D* X-ray diffractogram in the left and right panel of Fig. 1(b), respectively. It is also observed that initially, the intensity of the (104) peak was more than the (110) peak. Still, as the annealing temperature is raised, the intensity of both peaks becomes nearly equal. For sample $T_1$, both peaks are almost merged. Still, as the annealing temperature increases to 550 °C, the peaks can be resolved well but shifted slightly towards the higher angle side. Finally, at the annealing temperature of 600 °C, the peaks are resolved well and returned to their original positions. Further, to get more insight into structural information, Rietveld refinement of all XRD patterns is performed using the FullProf suite by adopting the Thomson-Cox-Hasting Pseudo voigt function and a 6$^{th}$-order polynomial fitting for the background. The Global parameters, such as scale factor, zero shift, and background parameters, are refined first. Later, lattice parameters are refined to eliminate any deviation in peak position. After that, peak shape parameters and *FWHM* parameters are refined to obtain a decent fit. All parameters are refined in 5-10 cycles until the value of $\chi^2$ (Goodness of fit) reaches close to unity. XRD patterns of all samples (except $T_5$) have been refined using the same process to determine all structural parameters. The sample $T_5$ could not be well refined by considering *R3c* only because of a considerable amount of other phases. The final refined results of all XRD patterns and Bragg positions are shown in Fig. 1(c). Fitted and observed data are represented with black solid lines and red circles, respectively. The vertical green dash below each peak represents the Bragg position of the *R3c* phase of BFO, and the blue line at the bottom represents the difference between fitted and observed data. The fitting quality assessment is best done by following the refinement's graphical results [29]. From Fig. 1(c), it can be observed that the simulated XRD pattern matched quite well with the experimentally measured one. Tables 1 and 2 summarize the various parameters obtained from refinement, such as unit cell parameters, density, atomic positions, and $\chi^2$. The refined unit cell of BFO for the *R3c* phase is drawn in Fig. 1(d).

Crystallite size is usually calculated from the broadening of the peaks. Multiple factors like crystallite size, strain, temperature, and instrumental factors cause this broadening. Scherrer's method considers broadening due to crystallite size only, whereas the Williamson-Hall plot method includes the effect of crystallite size and strain on the broadening of peaks. The crystallite size ($D$) is calculated using Scherrer's formula $D = \frac{k\lambda}{\beta \cos\theta}$, here, $k$ is the shape factor (usually between 0.89-0.94), $\lambda$ is the wavelength used in the XRD observation, $\beta$ is *FWHM* for a particular peak, and $\theta$ is the Bragg angle. We know nanoparticles may be strained due to their surrounding conditions [30]. This fact is also considered in the estimation of crystallite size. For that, the crystallite size is also calculated along with strain analysis using a Williamson-Hall (WH) and the *FWHM* values, as obtained by Gaussian fitting of the highest intensity peaks (104) and (110), are used for this purpose. The modified equation can be written by including the broadening effect due to the strain in Scherrer's formula. Therefore, the total broadening is $\beta = \beta_{size} + \beta_{strain}$. Now using the above-given Scherrer's expression, we can get

$$\beta \cos\theta = \frac{k\lambda}{D} + 4\eta \sin\theta \qquad (2)$$

Rietveld refined *FWHM* values can also be used to calculate the crystallite size [31, 32]. The *FWHM* in Rietveld analysis is calculated using the following relation.

$$(FWHM)^2 = (U + D_{ST}^2)\tan^2\theta + V\tan\theta + W + \frac{IG}{\cos^2\theta}. \qquad (3)$$

Where the *U, V, W,* and *IG* are *FWHM* parameters. $D_{ST}$ is a strain-related parameter. In the Rietveld refinement, these parameters are meant to be refined very carefully by considering the effect of instrumental broadening, too. The value of crystallite sizes obtained from all three methods is plotted in Fig. 2(a-d). It is found that the crystallite size increases with an increase in the annealing temperature. This is due to the growth of the crystal with increasing annealing temperature. A comparison between Scherrer's and WH plot methods shows that Scherer's method yields a low crystallite size compared to the latter. This may be because we usually neglect the broadening due to strain in the first. The strain analysis found that the strain decreases with increased annealing temperature due to dense and fine compact crystals forming, reducing the strain values [33]. The same reason can also be assumed to explain the increasing crystallite size, which may eventually decrease the strain in nanocrystals. The dislocation density ($\delta$) estimated as the length of dislocation lines per unit volume, is expressed as $\delta = 1/D^2$ in the unit of nm$^{-2}$. The $\delta$-values are calculated from crystallite sizes obtained from all three methods. It is found that the dislocation density of the crystallites decreases with increment in the annealing temperature. This also confirms the higher crystallinity of material at higher annealing temperatures. The estimated crystallite size, strain, and dislocation density values are plotted in Fig. (3) and tabulated in Table 2.

Bond length and bond angles can also be determined from the Rietveld analysis. It is found that initially, bond lengths decrease with an increase in the annealing temperature. At higher annealing temperatures, the bond length stabilized and remained unchanged. This can be explained by stabilizing the structure at a higher annealing temperature and reducing strain, as confirmed by WH analysis. In FeO$_6$ octahedra, the long bond between Fe and O decreases slightly and becomes stable at higher annealing temperatures. Similarly, the short bond between Fe and O shows an initial increment and stabilizes at a higher annealing temperature. A similar trend is also observed for the bonds in the Bi-O polyhedral. The Fe-O-Fe and O-Fe-O bond angles are increased initially and later stabilized at higher annealing temperatures. In contrast, the Bi-O-Bi bond angle decreases initially and is stable later. Hence, the bond length and bond angle become stable at higher annealing temperatures, confirming a stable crystal structure. Values of bond length and bond angles obtained from the analysis are shown in Table 3.

The characteristic factor of a typical perovskite structure, the Goldschmidt tolerance factor (*t*), which is used to indicate how much a given material deviates from the typical perovskite structure, is also calculated using in Eq. (1).

$$t = \frac{Bond\ length\ of\ Bi\_O\ bond}{\sqrt{2}(Bond\ length\ of\ Fe\_O\ bond)} \qquad (4)$$

In our case, the tolerance factor for BFO is found to be 0.89, as estimated from the values Bi-O$_2$ and Fe-O$_{short}$ given in Table 3. This value is also in the range of accepted values for rhombohedrally distorted perovskite [6]. The electron density plots represent the interactions between constituents at the atomic level. The electron density $\rho(x, y, z)$ can be estimated from the structure factor described below.

$$\rho_{hkl}(x, y, z) = \Sigma F_{hkl} \frac{e^{-2\pi i(hx+ky+lz)}}{V} \quad (5)$$

Here, $\rho(x, y, z)$ represents the electron density; $F_{hkl}$ is the structure factor for the respective Miller plane $(hkl)$; and $V$ is the unit cell volume. The cross-section of the structure in the $zy$ plane makes an intercept of $x = 0$ on the $x$-axis, and the $xy$ plane at $z = 0$ intercept and the electron density plots for BFO are demonstrated for these planes in Fig. 4. The variation in electron density at the atomic level is due to the structural changes which can eventually influence the multiferroic properties of BFO nanoparticles.

### 3.2. FTIR analysis

The FTIR spectra of BFO nanoparticles annealed at different temperatures are shown in Fig. 5(a). A sharp absorption band is observed in the 400-850 cm$^{-1}$ range, corresponding to metal-oxygen bonds existing in BFO, as shown in Fig. 5(b). Strong absorption peaks have been observed around 442 cm$^{-1}$ and 575 cm$^{-1}$ due to the stretching vibration of Fe-O bonds in FeO$_6$ octahedra. Another strong absorption peak is observed at 525 cm$^{-1}$ due to the stretching vibration of the Bi-O bonds and at 850 cm$^{-1}$ due to the bending vibration of Bi-O-Bi bonds. These bonds are the thumbprints of tetrahedral FeO$_4$ and octahedral FeO$_6$ groups in perovskite compounds [34]. In samples T$_4$ and T$_5$, an absorption peak is found around 815 cm$^{-1}$, which is due to the vibration of Fe-O bonds in the FeO$_4$ tetrahedra that exist in the impurity phases like Bi$_2$Fe$_4$O$_9$ and Bi$_{25}$FeO$_{40}$. Besides these, a broad absorption band is found in the 1250-1700 cm$^{-1}$ range, mainly attributed to some remaining organic molecules in the sample. A peak at 1384 cm$^{-1}$ corresponds to trapped nitrate ions in BFO. Another considerable peak is observed at 1740 cm$^{-1}$, corresponding to the –C=O stretching vibration in tartaric acid used in the synthesis [35]. A sharp but narrow absorption peak appeared at 2330 cm$^{-1}$ due to the O=C=O stretching vibration in the CO$_2$ molecule. A broad absorption band centred at 3000 cm$^{-1}$ is due to tartaric acid's –OH stretching vibration. Another wide absorption band is also found at 3600 cm$^{-1}$ due to water molecules. Table 4 summarizes the list of chemical bonds and their group identified from the FTIR spectroscopy. Hence, the FTIR analysis mainly confirms the existence of Fe-O and Bi-O bonds in the BFO nanoparticles.

### 3.3. UV-Visible analysis

The UV-visible absorbance spectra of BFO nanoparticles annealed at different temperatures are shown in Fig. 6(a). A strong response can be seen in the visible region below 600 nm. The threshold edge can be defined near this wavelength to identify the bandgap. Tauc's relation $\alpha E = A(E - E_g)^q$ is used to estimate the band gap [36]. $E_g$ is the material's bandgap, $A$ is a constant, $E$ is the incident photon's energy, and $q$ is a constant number used for a particular transition. $Q$ can be 1/2 or 2 for direct or indirect electronic transitions. Graph of $(\alpha E)^2$ and $E$ is plotted, and on the extrapolation of the linear portion of the curve, which intersects on the *x*-axis, gives the bandgap value. Tauc's plot for all samples is shown in Fig. 6(b), and corresponding estimated bandgap values are plotted in Fig 6(c). It has been observed that the value of the bandgap slightly decreases with an increase in the annealing temperature. This is due to the quantum size effect caused by an increment in the crystallite size and the annealing temperature [37], as discussed earlier in the XRD analysis section. Hence, the UV-visible analysis of BFO nanoparticles confirms its response in the UV and visible range and the bandgap of this material to be in the visible region. Therefore, we can further explore the application of this material in photo absorption-related phenomena like photovoltaics and photocatalysis [38].

### 3.4. Thermal analysis

Thermal analysis of the crystalline BFO nanoparticles is studied using differential scanning calorimetry (DSC). Figure 7 shows the DSC thermograms obtained for synthesized BFO nanoparticles in the temperature range of 40 °C – 400 °C with a heating rate of 10 °C/min. In every thermogram at the beginning around 50 °C, a sharp endothermic start-up hook can be seen, which may be attributed to variations in the sample's heat capacity and reference, melting and evaporation of remanent organic impurities, and removal of moisture content. Another activity near 200 °C is observed for $T_1$ and $T_2$ samples, possibly due to the change in specific heat and reduction in the amorphousity [39]. Apart from these two, no other significant thermal activity is detected. This confirms the thermal stability of crystalline material in this temperature range. From the thermal analysis and structural analysis (performed earlier), it is clear that the sample annealed at 550 °C is the most stable with the highest crystallinity.

### 3.5. VSM Analysis

The magnetic properties of BFO nanoparticles at room temperature are studied using the VSM technique. The magnetic hysteresis (*M-H* loop) of all samples is investigated over the magnetic field in the range of ± 10,000 Gauss, as shown in Fig. 8(a). The *M-H* loops confirm the weak ferromagnetic nature of BFO nanoparticles. The $Fe^{+3}$ and some oxygen vacancies cause the magnetic character of BFO nanoparticles. Since the magnetic ordering in BFO is presumably antiferromagnetic, *M-H* loops could not be saturated even at the higher magnetic fields. *M-H* curves obtained from the VSM are fitted and analyzed using Eq. (6) to separate the ferromagnetic and antiferromagnetic contributions.

$$M(H) = \left(\frac{2M_{FM}^S}{\pi}\right) \tan^{-1}\left[\left(\frac{H \pm H_c}{H_c}\right) \tan\left\{\frac{\pi M_{FM}^R}{2M_{FM}^S}\right\}\right] + \chi H \qquad (6)$$

The fitted *M-H* curves for different BFO nanoparticles are shown in Fig. 8(b-f). This provides the contribution of the ferromagnetic and antiferromagnetic part and calculated parameters such as saturation ($M_{FM}^S$) and remnant ($M_{FM}^R$) magnetization, coercivity ($H_C$), and antiferromagnetic susceptibility ($\chi$). These parameters are also tabulated in Table 5. The magnetization values increase with the annealing temperature till 550 °C and then start to decrease due to a substantial proportion of non-magnetic phases like $Bi_2Fe_4O_9$ and $Bi_{25}FeO_{40}$ [40]. A maximum value of $M_S \sim 0.54$ emu/g is obtained for sample $T_3$. This also ensures that this sample has the highest crystallinity, as earlier confirmed by the structural and thermal analysis. It is known that particle size and specific surface area of nanoparticles can severely affect magnetization [42]. Nanoparticles usually have a higher specific surface area, so the uncompensated surface spins become independent of spin arrangement, which is the main reason for the observed magnetization in BFO nanoparticles. Usually, BFO is found to be antiferromagnetic in bulk form. This weak ferromagnetism in these nanoparticles is caused by suppressing the spin cycloid, the oxygen vacancies, and the fluctuation in the valence of $Fe^{+3}$ and $Fe^{+2}$ cations [43]. For the case of bulk BFO, a spiral spin structure with a period length of around 62 nm is the main reason for the suppression of magnetization [43, 44]. However, the crystallite sizes of these nanoparticles, as estimated by XRD, are less than the cycloid spin period; BFO nanoparticles exhibit weak ferromagnetism compared to the complete antiferromagnetism found in bulk BFO [43].

## 5. Conclusion

Using an efficient sol-gel method, we have successfully synthesized the single-phase BFO nanoparticles at lower annealing temperatures. X-ray diffraction has confirmed the crystalline nature of these nanoparticles. Single-phase Rietveld refinement of diffraction patterns has ensured the rhombohedral structure with the *R3c* space group symmetry. Some impurity phases like $Bi_2Fe_4O_9$ and $Bi_{25}FeO_{40}$ are observed in samples annealed at 600 °C and 650 °C. The crystallite size is estimated using Scherrer's formula, the Williamson-Hall plot method, and Rietveld analysis. Crystallite sizes are found to increase with the annealing temperature. Dislocation densities and strain decrease with an increase in the annealing temperature. Bond lengths and bond angles decrease initially and then stabilize at a higher annealing temperature, suggesting the formation of a stable crystal structure. From the UV-visible spectra analysis, a strong response below 600 nm is observed, ensuring that the bandgap lies in the optical region, which decreases with an increase in the annealing temperature. No significant thermal activity is observed from DSC analysis, confirming the thermal stability of the synthesized BFO nanoparticles. The ferromagnetic part of total magnetization first increases with annealing temperature and then decreases due to the formation of other non-magnetic phases. A maximum value of saturated magnetic moment around 0.54 emu/g is estimated by fitting a magnetic hysteresis loop for the nanoparticles synthesized at 550 °C. Thus, this study confirms that it is possible to synthesize single-phase BFO nanoparticles at lower annealing temperatures with a bandgap in the optical regime, making these synthesized BFO nanoparticles suitable for their applications in photocatalysis and gas-sensing.

## Acknowledgement

This research is supported by the research grant from the Science and Engineering Research Board (SERB), Govt. of India, against scheme ECR/2017/001612. HP also acknowledges the SVNIT Institute seed Money grant 2021-22/DOP/04. The authors are thankful to Ms Aditi for UV-visible measurements and to the Department of Science and Technology (DST), Govt. of India, for the infrastructural support through the FIST grant, sanction number SR/FST/PS-I/2017/12 for XRD measurements.

## Conflict of Interest

The authors declare no conflict of interest regarding this article.

## Data Availability Statement

The data supporting this study's findings are available from the corresponding author upon reasonable request.

## References
[1] SW. Cheong, M. Mostovoy: Nat. Mater. 6(1) (2007) 13. DOI: https://doi.org/10.1038/nmat1804.
[2] JF. Scott: Nat. Mater. 6 (2007) 256. DOI: https://doi.org/10.1038/nmat1868.
[3] S. Seki, X.Z. Yu, S. Ishiwata, Y. Tokura: Science 336(6078) (2012) 198. DOI: https://doi.org/10.1126/science.1214143.

[4] H. Schmid: ferroelectrics 162(1) (1994) 317. DOI: https://doi.org/10.1080/00150199408245120.
[5] N.A. Hill: J. Phys. Chem. 104(29) (2000) 6694. DOI: 10.1021/jp000114x.
[6] G. Catalan, J. Scott: Adv. Mater. 21(24) (2009) 2463. DOI: doi.org/10.1002/adma.200802849.
[7] J. Yang, Q. He, P. Yu Y. Chu: Ann. Rev. Mater. Res. 45 (2015) 249. DOI: 10.1146/annurev-matsci-070214-020837.
[8] H. Megaw, C. Darlington: Acta. Crystallogr. Scet. A 31(2) (1975) 161. DOI: 10.1107/S0567739475000332.
[9] Wang J, Neaton J. B, Zheng H, Nagarajan V, Ogale SB, Liu B, Viehland D, Vaithyanathan V, Schlom DG, Waghmare UV, Spaldin NA, Rabe K. M, Wuttig M: Sci. 299 (2003), 1719. DOI: 10.1126/science.1080615.
[10] M. Sakar, S. Balakumar, P. Saravanan, S.N. Jaisankar: Mater. Res. Bull. 48(8) (2013) 2878. DOI: 10.1016/j.materresbull.2013.04.008.
[11] T. Pikula, T. Szumiata, K. Siedliska, V.I. Mitsiuk, R. Panek, M. Kowalczyk, E. Jartych: Metall. Mater. Trans. A 53 (2022) 470. DOI: https://doi.org/10.1007/s11661-021-06506-z.
[12] Singh, H; Rajput, JK: SN Appl. Sci. 20(2) (2020) 31. https://doi.org/10.1007/s42452-020-3140-2.
[13] G. Marschick, J. Schell, B. Stöger, J. N. Gonçalves, M. O. Karabasov, D. Zyabkin, A. Welker, M. Escobar C., D. Gärtner, I. Efe, R. A. Santos, J. E. M. Laulainen, and D. C. Lupascu, Phys. Rev. B 102(22) (2020) 224110. DOI: 10.1103/PhysRevB.102.224110.
[14] Jian-Ping Zhou, Ruo-Lin Yang, Rui-Juan Xiao, Xiao-Ming Chen, Chao-Yong Deng: Mater. Res. Bull. 47(11) (2012) 3630. DOI: https://doi.org/10.1016/j.materresbull.2012.06.050.
[15] Wani W.A., Kundu S., Ramaswamy K: SN Appl. Sci. 2 (2020) 1969. DOI: 10.1007/s42452-020-03669-z.
[16] Wu, Jiagang, Zhen Fan, Ding-quan Xiao, Jianguo Zhu, John Wang: Prog. Mater. Sci. 84 (2016) 335. DOI: 10.1016/J.PMATSCI.2016.09.001.
[17] Goldschmidt, V. M: Trans. Faraday Soc. 25 (1929) 253. DOI : 10.1039/TF9292500253.
[18] Arti, Sumit Kumar, Parveen Kumar, Rajan Walia, Vivek Verma: Results Phys. 14 (2019) 102403. DOI: 10.1016/j.rinp.2019.102403.
[19] M Kumar, M Arora, S Chauhan, H Pandey: Journal of Alloys and Compounds 735 (2018) 684-691. DOI: 10.1016/j.jallcom.2017.11.152
[20] Zhang, Xinyi, Lai, Chin wei, Zhao, Xiangyong Wang, Da Yu Dai, Jiang: Appl. Phys. Lett. 87 (2005) 143102. DOI: 10.1063/1.2076437.
[21] S Chauhan, M Kumar, H Pandey, S Chhoker: SC Katyal, Journal of Alloys and Compounds 811 (2019) 151965. DOI: 10.1016/j.jallcom.2019.151965
[22] S Joshi, M Kumar, H Pandey, M Singh, P Pal: Journal of Alloys and Compounds 768 (2018) 287-297. DOI: 10.1016/j.jallcom.2018.07.250
[23] Scott J., Morrison Finlay, Miyake M., Zubko P: Ferroelectrics 336 (2006) 237. DOI: 10.1080/00150190600697699.
[24] S Joshi, M Kumar, H Pandey, S Chhoker: AIP Conference Proceedings 2136 (2019) 040007. DOI: 10.1063/1.5120921
[25] Rani Sonu, Shekhar Mukesh, Kumar Pawan, Prasad, Surabhi: Appl. Phys. A 128(12) (2022) 1046. DOI: 10.1007/s00339-022-06171-y.

[26] M. Kumar and H. Pandey: Journal of Superconductivity and Novel Magnetism 36 (2023) 1269–1276. DOI: 10.1007/s10948-023-06568-7

[27] Haumont R., Kornev Igor, Lisenkov S., Bellaiche L., Kreisel, J., Dkhil Brahim: Phys. Rev. B. 78 (2008) 13. DOI: 10.1103/PhysRevB.78.134108.

[28] D. Karoblis, D. Griesiute, K. Mazeika, D. Baltrunas, D. V. Karpinsky, A.Lukowiak, Pawel Gluchowski, R. Raudonis, A. Katelnikovas, A. Zarkov: Materials 13(13) (2020) 3035. DOI: 10.3390/ma13133035.

[29] B. Toby: Powder Diffr. 21(1) (2006) 67-70. DOI: 10.1154/1.2179804.

[30] G. Ouyang, W.G. Zhu, C.Q. Sun, Z.M. Zhu, S.Z. Liao: Phys. Chem. Chem. Phys. 12 (2010) 1543. DOI: https://doi.org/10.1039/B919982A.

[31] M. Karolus, E. Lagiewka: J. Alloys Compd. 367 (2004) 235. DOI: 10.1016/j.jallcom.2003.08.044.

[32] L. Lutterotti, P. Scardi: J. Appl. Cryst. 23 (1990) 246. DOI: 10.1107/S0021889890002382.

[33] V. Ramasamy, Y. Subramanian, S. Varadarajan, K. Ramaswamy, K. Kaliappan, D. Arulmozhi, G. R. Srinivasan, R. K. Gubendiran: Ceram. Int. 46(2) (2020) 1457. DOI: 10.1016/j.ceramint.2019.09.111.

[34] P. Hermet, M. Goffinet, J. Kreisel, P.H. Ghosez: Phys. Rev. B 75 (2007) 220102. DOI: 10.1103/PhysRevB.75.220102.

[35] P.M. Chu, F.R. Guenther, G.C. Rhoderick, W.J. Lafferty: J. Res. Natl. Inst. Stand. Technol. 104 (1999) 59. DOI: https://dx.doi.org/10.6028/jres.104.004.

[36] Tauc, J: Mater. Res. Bull. 3 (1968) 37. DOI: https://doi.org/10.1016/0025-5408(68)90023-8.

[37] M. Singh: ICMETE (2018) 339. DOI: https://doi.org/10.1109/ICMETE.2018.00080.

[38] Gulati Shikha, Goyal Kartika, Arora Aryan, Kumar Sanjay, Trivedi Manoj, Jain Shradha: Environ. Sci.: Water Res. Technol 8 (2022) 1590. DOI: 10.1039/D2EW00027J.

[39] J.H. Xu, H. Ke, D.C. Jia, W. Wang, Y. Zhou: J. Alloy. Compd. 472 (2009) 473. DOI: https://doi.org/10.1016/j.jallcom.2008.04.090.

[40] P. Sharma, A. Kumar, J. Tang, G. Tan: Magnetic Materials and Magnetic Levitation, IntechOpen (2021). DOI: 10.5772/intechopen.93280.

[41] M.B. Stearns, Y. Cheng: J. Appl. Phys. 75 (1994) 6894. DOI: https://doi.org/10.1063/1.356773.

[42] S. Mohammadi, H. Sholrollahi, M. Basiri: J. Magn. Magn. Mater. 375 (2015) 38. DOI: https://doi.org/10.1016/j.jmmm.2014.09.050.

[43] I. Sonowska, T. Pterlin-Neumaier, E. Steichele: J. Phys. C 15 (1982) 4835. DOI: 10.1088/0022-3719/15/23/020.

[44] J. Wang, JB. Neaton, H. Zheng, V. Nagarajan, SB. Ogale, B. Liu, D. Viehland, V. Vaithyanathan, D.G. Scholm, U.V. Waghmare, N.A. Spaldin, K.M. Rabe, M. Wuttig, R. Ramesh: Science 299(5613) (2003) 1719. DOI: https://doi.org/10.1126/science.1080615.

**List of Figures**

**Figure 1:** (a) XRD profiles of BFO nanoparticles annealed at different temperatures (b) variation in position and intensity of significant peaks (012) and (104); (110) with annealing temperature (c) Rietveld refined XRD profiles for *R3c* phase (d) Crystal structure of BFO.

**Figure 2:** Williamson-Hall plots to estimate the crystallite size and strain.

**Figure 3:** Variation of (a) crystallite sizes, (b) dislocation density, and (c) strain, as obtained using different methods, with the annealing temperature.

**Figure 4:** Electron density distribution estimated from Rietveld analysis for (a) *xy* plane and (b) *yz* plane, respectively, for BFO nanoparticles annealed at 550 °C.

**Figure 5:** (a) IR transmittance spectra of BFO nanoparticles in the range of 400-4000 cm$^{-1}$ (b) same spectra in the range of 400 – 850 cm$^{-1}$ for better clarity.

**Figure 6:** (a) UV Visible transmittance spectra of BFO nanoparticles in the range of 190-800 nm (b) Estimation of band gap using Tauc's plot (c) Variation in the band gap with annealing temperature.

**Figure 7:** DSC thermograms of BFO nanoparticles measured in 40-400 °C range.

**Figure 8:** (a) The room temperature magnetic hysteresis loop for BFO nanoparticles annealed at different temperatures (b-f) The MH curve fitted using Eq. (6) to separate the ferromagnetic and antiferromagnetic contributions.

**List of tables**

**Table 1:** Lattice parameters and atomic positions obtained from Rietveld refinement for the *R3c* phase.

**Table 2:** Comparison between crystallite sizes (*D*), dislocation densities (*δ*), and strain (*η*) values obtained using different methods.

**Table 3:** Bond lengths and bond angles obtained from the Rietveld analysis.

**Table 4:** List of chemical bonds and their groups identified from the FTIR spectroscopy.

**Table 5:** Magnetic parameters obtained from the fitting of the MH loops.

**Figure 1**

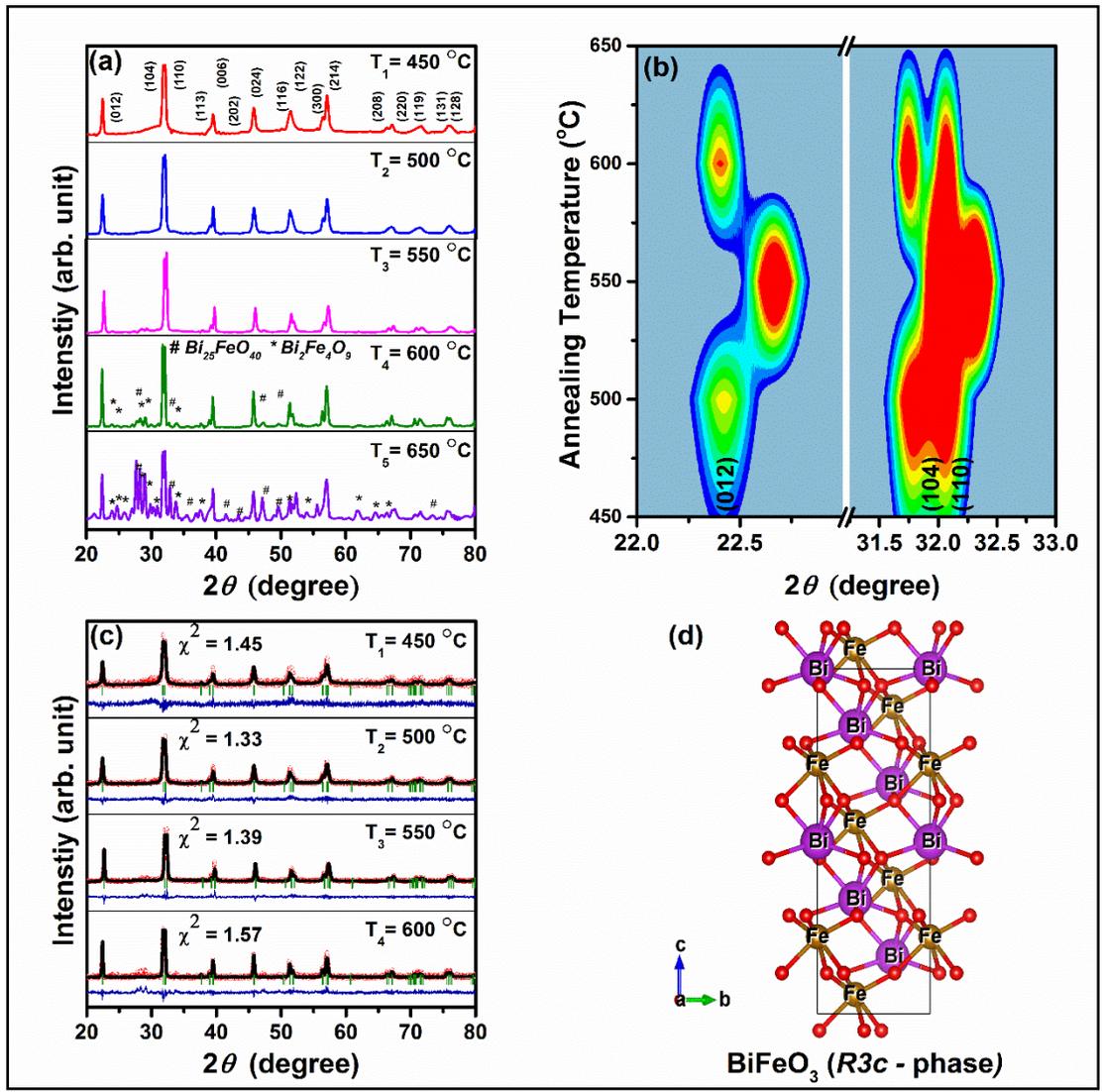

**Figure 2**

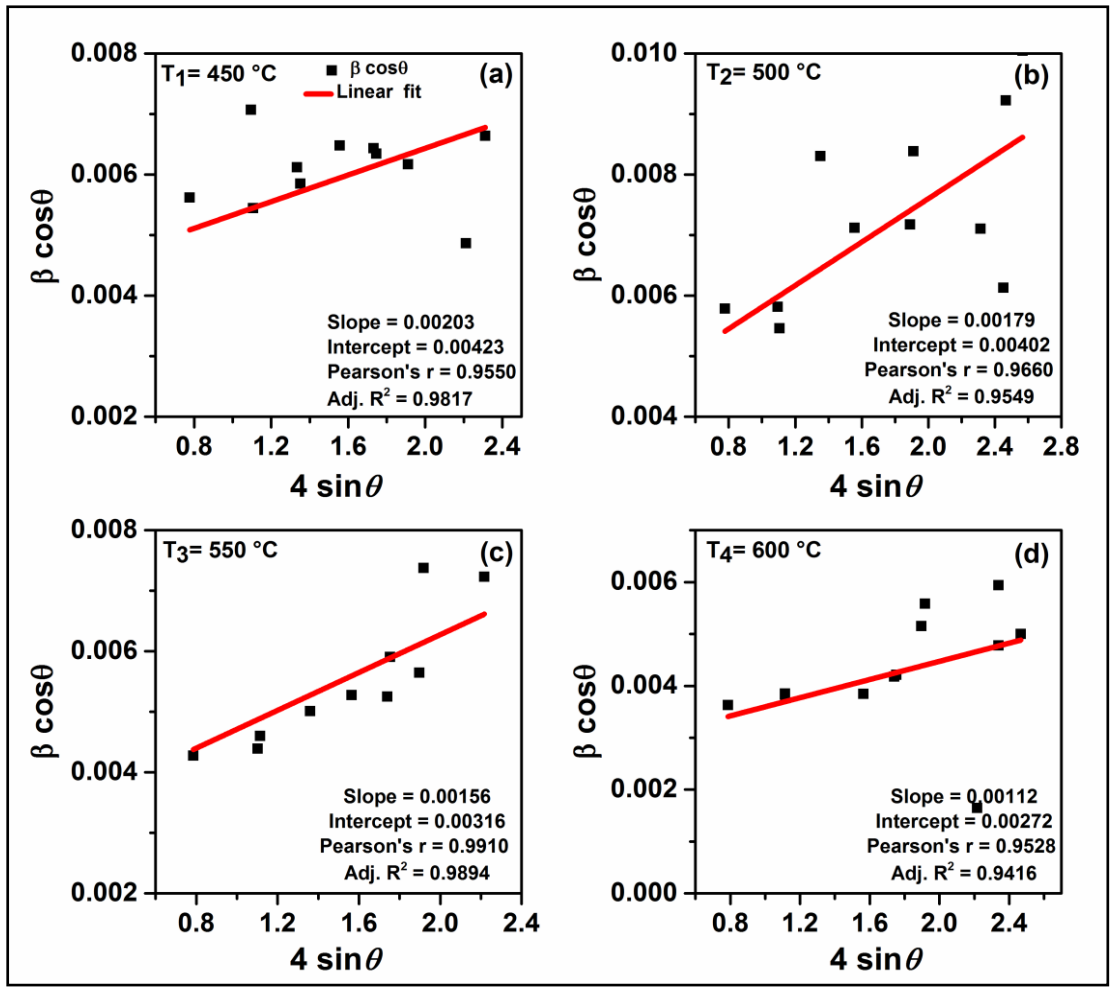

**Figure 3**

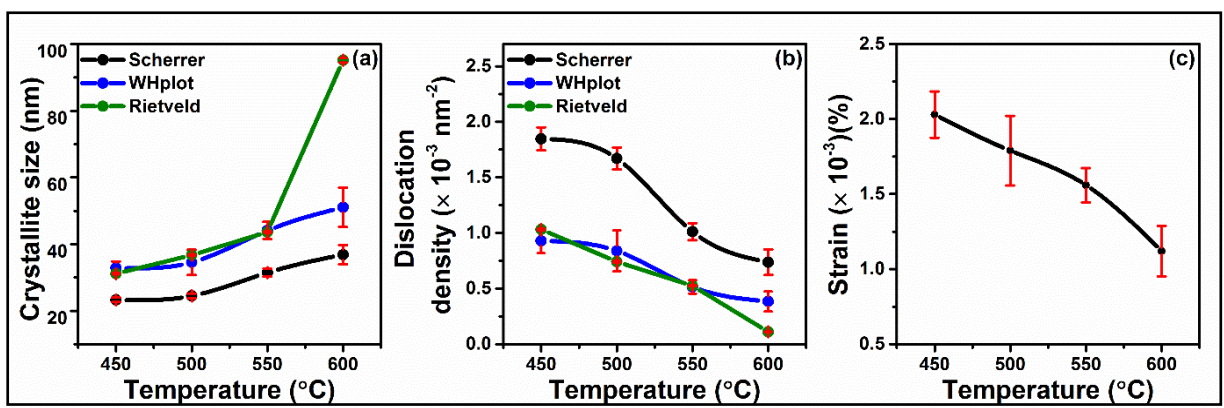

**Figure 4**

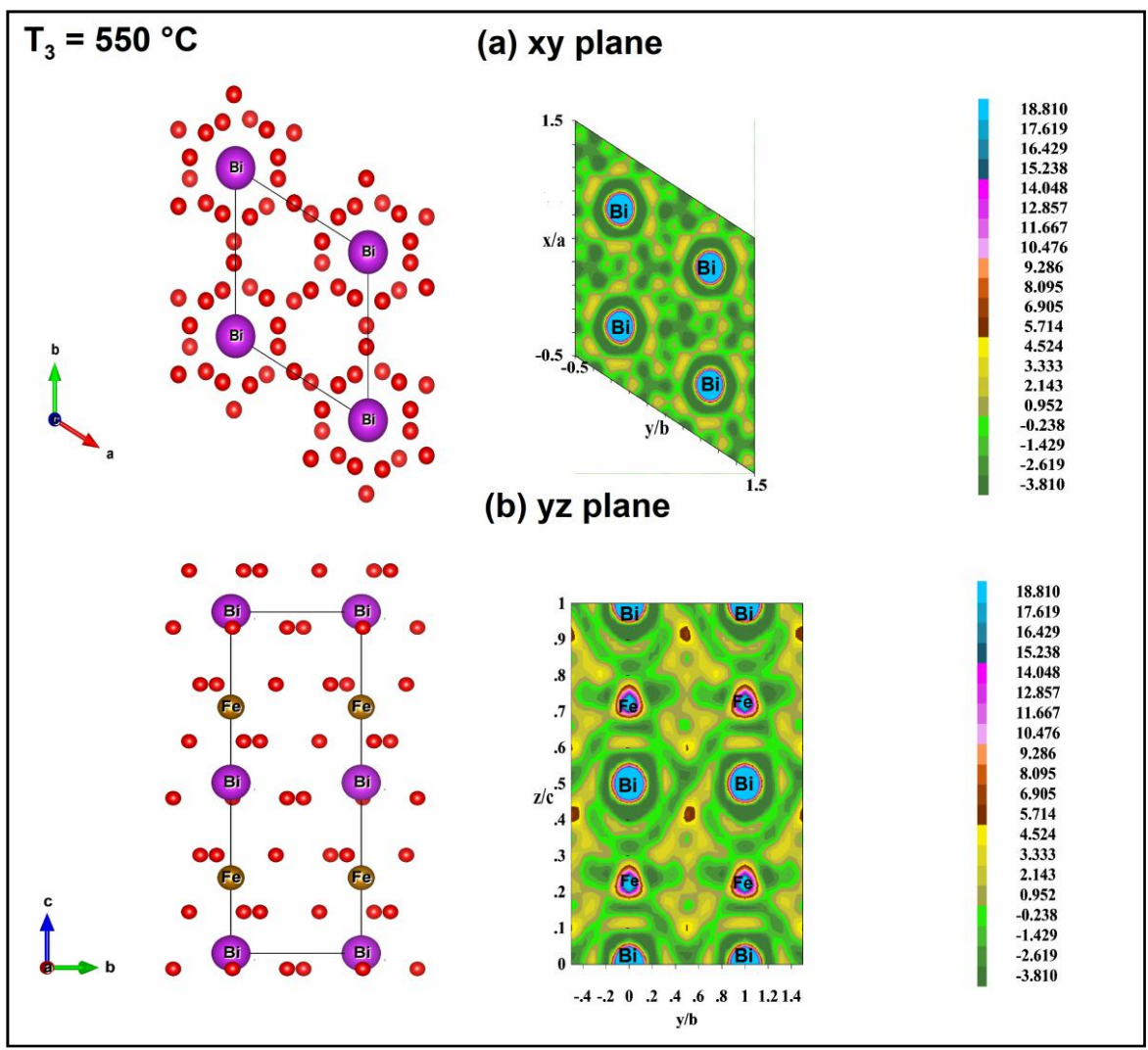

**Figure 5**

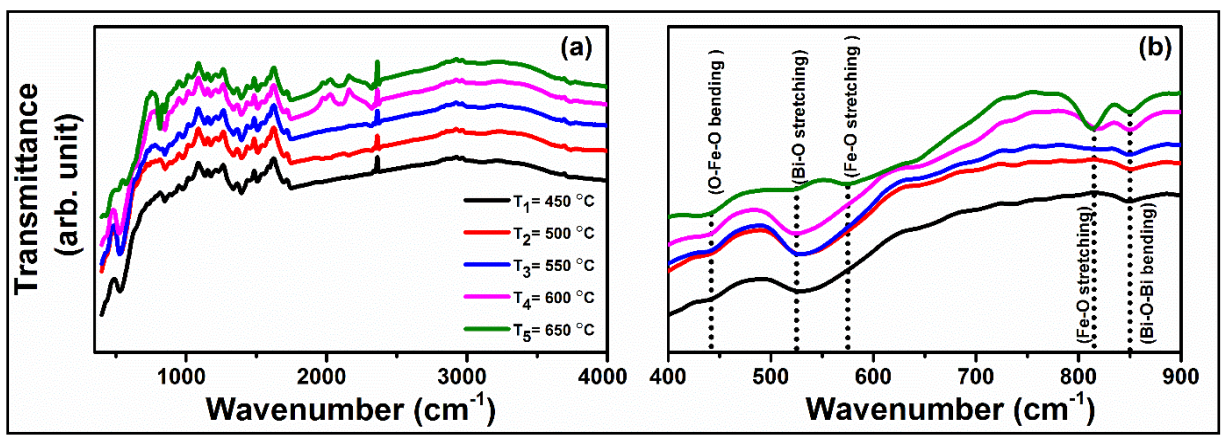

**Figure 6**

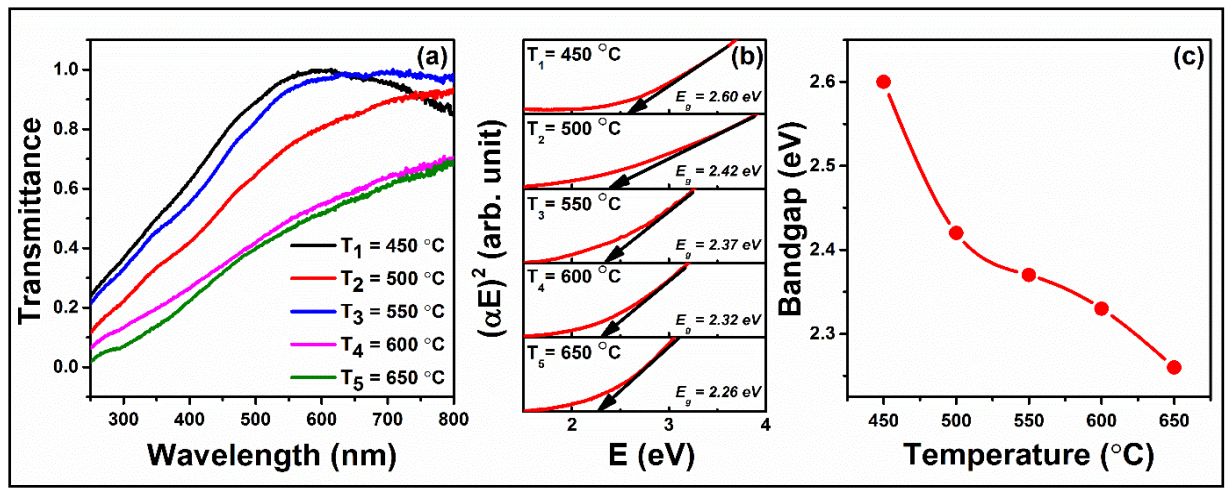

**Figure 7**

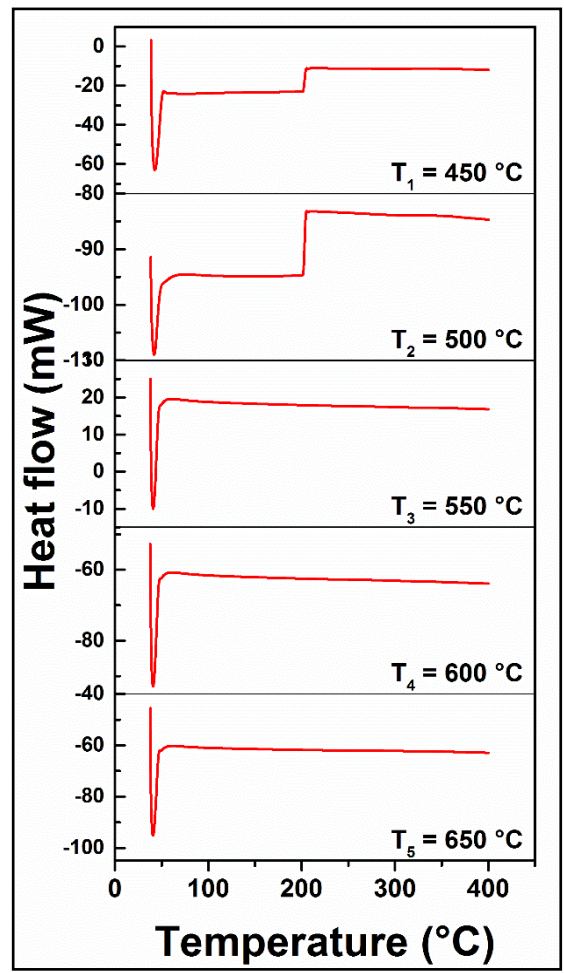

**Figure 8**

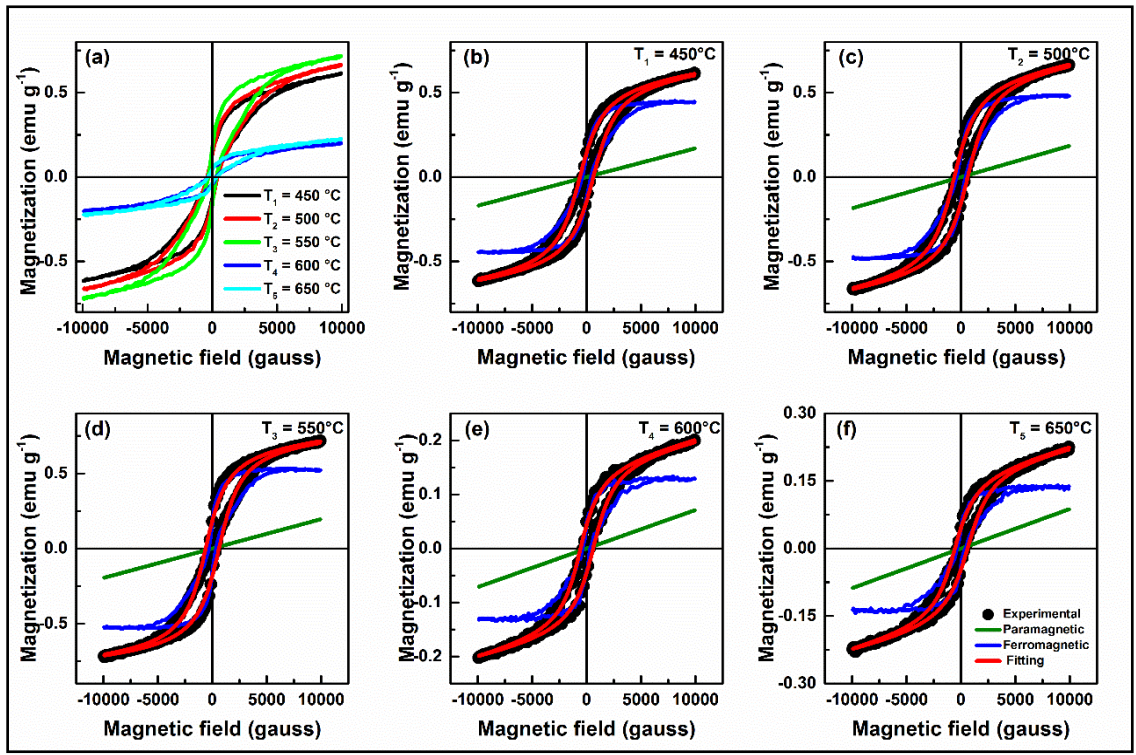

# Table 1

| Annealing Temperature (°C) | Lattice parameters | | | Atomic positions | | | | $\chi^2$-value |
|---|---|---|---|---|---|---|---|---|
| | $a=b$ (Å) | $c$ (Å) | Density $\rho$ (g cm$^{-3}$) | Atom | $x$ | $y$ | $z$ | |
| 450 | 5.5700 | 13.8482 | 8.380 | Fe | 0 | 0 | 0.2201 | 1.42 |
| | | | | Bi | 0 | 0 | 0.4978 | |
| | | | | O | 0.1043 | 0.2968 | 0.2846 | |
| 500 | 5.5756 | 13.8575 | 8.354 | Fe | 0 | 0 | 0.2179 | 1.33 |
| | | | | Bi | 0 | 0 | 0.4963 | |
| | | | | O | 0.0960 | 0.3108 | 0.2883 | |
| 550 | 5.5781 | 13.8676 | 8.341 | Fe | 0 | 0 | 0.2199 | 1.39 |
| | | | | Bi | 0 | 0 | 0.4974 | |
| | | | | O | 0.0893 | 0.2956 | 0.2886 | |
| 600 | 5.5773 | 13.8672 | 8.343 | Fe | 0 | 0 | 0.2273 | 1.57 |
| | | | | Bi | 0 | 0 | 0.5059 | |
| | | | | O | 0.0967 | 0.3433 | 0.2976 | |

# Table 2

| Annealing Temperature | | $T_1 = 450$ °C | $T_2 = 500$ °C | $T_3 = 550$ °C | $T_4 = 600$ °C |
|---|---|---|---|---|---|
| Scherer's method | $D$ (nm) | 23.27 ± 0.65 | 24.46 ± 0.73 | 31.43 ± 1.19 | 36.84 ± 2.87 |
| | $\delta \times 10^{-3}$ (nm$^{-2}$) | 1.84 ± 0.10 | 1.67 ± 0.09 | 1.01 ± 0.07 | 0.73 ± 0.11 |
| WH Plot method | $D$ (nm) | 32.82 ± 1.97 | 34.53 ± 3.79 | 44.07 ± 2.58 | 51.04 ± 5.93 |
| | $\delta \times 10^{-3}$ (nm$^{-2}$) | 0.92 ± 0.11 | 0.83 ± 0.18 | 0.51 ± 0.060 | 0.384 ± 0.08 |
| | $\eta \times 10^{-3}$ (%) | 2.03 ± 0.15 | 1.79 ± 0.23 | 1.59 ± 0.11 | 1.12 ± 0.17 |
| Rietveld method | $D$ (nm) | 31.13 ± 0.16 | 36.73 ± 0.25 | 43.64 ± 0.18 | 95.14 ± 0.71 |
| | $\delta \times 10^{-3}$ (nm$^{-2}$) | 1.03 ± 0.01 | 0.741 ± 0.01 | 0.52 ± 0.01 | 0.11 ± 0.01 |

## Table 3

|  |  | $T_1 = 450\ °C$ | $T_2 = 500\ °C$ | $T_3 = 550\ °C$ | $T_4 = 600\ °C$ |
|---|---|---|---|---|---|
| Bond length (Å) | Fe-O$_{short}$ | 1.9224 | 1.9447 | 1.9454 | 1.9452 |
|  | Fe-O$_{long}$ | 2.1803 | 2.1493 | 2.1505 | 2.1504 |
|  | Bi-O$_1$ | 2.2619 | 2.2869 | 2.2882 | 2.2881 |
|  | Bi-O$_2$ | 2.4615 | 2.4568 | 2.4579 | 2.4577 |
|  | Bi-Fe | 3.1218 | 3.1093 | 3.1116 | 3.1117 |
| Bond angle (degree) | O-Fe-O | 161.1027 | 163.3079 | 163.3060 | 163.3051 |
|  | Fe-O-Fe | 149.382 | 150.7744 | 150.7754 | 150.7760 |
|  | O-Bi-O | 147.437 | 146.2378 | 146.2341 | 146.2326 |

## Table 4

| Bond | Type of vibration | Group | Wavenumber (cm$^{-1}$) |
|---|---|---|---|
| O-Fe-O | Bending | FeO$_6$ octahedra | 442 |
| Fe-O | Stretching | FeO$_6$ octahedra | 575 |
| Bi-O | Stretching | BiO polyhedra | 525 |
| Bi-O-Bi | Bending | BiO polyhedra | 850 |
| Fe-O | Stretching | FeO$_4$ tetrahedra | 815 |
| -C=O | Stretching | -COOH | ~ 1740 |
| O=C=O | Stretching | CO$_2$ | ~ 2330 |
| -O-H | Stretching | -OH | ~ 3000 |

## Table 5

| Annealing Temperature (°C) | Ferromagnetic contribution | | | Paramagnetic Susceptibility $\chi\ (\times 10^{-5})$ |
|---|---|---|---|---|
|  | Saturation magnetization $M_S$ (emu g$^{-1}$) | Retentivity $M_R$ (emu g$^{-1}$) | Coercivity $H_C$ (gauss) |  |
| 450 | 0.443 | 0.144 | 402 | 1.699 |
| 500 | 0.482 | 0.143 | 418 | 1.851 |
| 550 | 0.535 | 0.172 | 342 | 1.960 |
| 600 | 0.130 | 0.047 | 357 | 0.714 |
| 650 | 0.137 | 0.046 | 393 | 0.879 |